\documentclass[journal]{IEEEtran}

\usepackage{graphicx}
\usepackage{float}
\usepackage{dirtytalk}
\usepackage{amsmath}
\usepackage{tikz}
\usepackage{tabularx}
\usepackage{blindtext}
\usepackage{hyperref}
\usepackage{cite}
\usepackage{stfloats} 
\usepackage{algorithm} 
\usepackage{algpseudocode} 
\usepackage{array}
\usepackage{multirow}
\usepackage{footmisc}
\usepackage{footnote}
\begin{document}

\title{Joint Instantaneous Amplitude-Frequency Analysis of Vibration Signals for Vibration-Based Condition Monitoring of Rolling Bearings}

\author{Sulaiman~Aburakhia,~\IEEEmembership{Member,~IEEE,} 
        Ismail Hamieh,~\IEEEmembership{Member,~IEEE,} 
       and~Abdallah~Shami,~\IEEEmembership{Senior~Member,~IEEE}

\thanks{Sulaiman Aburakhia, Ismail Hamieh, and Abdallah Shami are with the Department of
Electrical and Computer Engineering, Western University, N6A 3K7, Canada (e-mail: saburakh@uwo.ca; abdallah.shami@uwo.ca).}}

\markboth{}%
{Shell \MakeLowercase{\textit{et al.}}: Bare Demo of IEEEtran.cls for IEEE Journals}

\maketitle

\begin{abstract}
Vibrations of damaged bearings are manifested as modulations in the amplitude of the generated vibration signal, making envelope analysis an effective approach for discriminating between healthy and abnormal vibration patterns. Motivated by this, we introduce a low-complexity method for vibration-based condition monitoring (VBCM) of rolling bearings based on envelope analysis. In the proposed method, the instantaneous amplitude (envelope) and instantaneous frequency of the vibration signal are jointly utilized to facilitate three novel envelope representations: instantaneous amplitude-frequency mapping (IAFM), instantaneous amplitude-frequency correlation (IAFC), and instantaneous energy-frequency distribution (IEFD). Maintaining temporal information, these representations effectively capture energy-frequency variations that are unique to the condition of the bearing, thereby enabling the extraction of discriminative features with high sensitivity to variations in operational conditions. Accordingly, six new highly discriminative features are engineered from these representations, capturing and characterizing their shapes. The experimental results show outstanding performance in detecting and diagnosing various fault types, demonstrating the effectiveness of the proposed method in capturing unique variations in energy and frequency between healthy and faulty bearings. Moreover, the proposed method has moderate computational complexity, meeting the requirements of real-time applications. Further, the Python code of the proposed method is made public to support collaborative research efforts and ensure the reproducibility of the presented work.
\end{abstract}


\IEEEpeerreviewmaketitle

\section{Introduction}
\IEEEPARstart{V}{ibeation}-based condition monitoring (VBCM) can be defined as a signal-based methodology for assessing a system condition based on its inherent vibration patterns. By monitoring changes in vibration signatures, which reflect a change in the system's current state, VBCM provides a non-invasive, real-time approach to continuously monitor the system's condition. The main advantages of VBCM  over alternative forms of condition monitoring include \cite{hos20}\cite{ran11}:
\begin{itemize}
    \item  Vibration sensors are non-intrusive and can be contactless, facilitating non-destructive condition monitoring.
    \item  Real-time acquisition of vibration signals can be conducted in situ, allowing for online local condition monitoring.
    \item Trending vibration analysis can be utilized to identify relevant conditions and conduct comparative analysis across diverse conditions or objects.
   \item  Vibration sensors are cost-effective and widely available, offering various specifications to suit various requirements.
   \item Vibration waveform responds instantly to changes in the monitored condition and, therefore, is suitable for continuous and intermittent monitoring applications. 
   \item Signal processing techniques can be applied to vibration signals to mitigate corrupting noise and extract weak condition indications from other masking signals.
\end{itemize}

The process of VBCM involves two main aspects: feature extraction and condition monitoring. Feature extraction involves analyzing vibration signatures to extract relevant features that reflect system condition changes. Such signatures often manifest through sudden changes in amplitude, frequency, and phase characteristics of the generated vibration pattern. Existing feature extraction methods include time-domain, frequency-domain, and time-frequency-domain methods. In time-domain-based methods \cite{td0, td1, td2, td4, td9, td15, td_en2, td_en3}, features are calculated from the signal's amplitude, representing specific characteristics of the signal's dynamics over its time period. Common types of time-domain features include shape features and statistical features. Shape features involve values such as maximum, minimum, peak-to-peak, and crest factor. Statistical features describe characteristics of the probability distribution of the signal, such as mean, standard deviation,  variance, skewness, and kurtosis. Time-domain feature extraction approaches are generally simple to implement and are computationally efficient since signal transformation is not required, making them advantageous in applications where real-time processing is crucial, especially with limited computational resources. However, time-domain analysis is highly susceptible to noise, which can mask the dynamic characteristics of the signal. Furthermore, to achieve reliable performance, input vibration segments of relatively large durations are usually required to precisely capture evolving changes and complexities within the segment. Frequency-based methods \cite{Attoui1, fd5, fd12, fd13, psd1, psd2, psd3, psd4} represent the signal in terms of its spectral contents, revealing details that are not apparent in the time-domain waveform. Accordingly, discriminative spectral features can be extracted from the signals' spectrum. In contrast to time-domain analysis, frequency-based analysis allows for identifying and removing noise or unwanted components by applying appropriate frequency filtering mechanisms. In the context of VBCM of rolling bearings,  the generated vibration pattern spans a broad spectrum of frequency components, including characteristic frequencies related to the bearing geometry and operational speed, as well as harmonic and sideband frequencies caused by various operational conditions. This makes spectral analysis particularly effective in discriminating between healthy and abnormal vibration patterns. However, in frequency-domain methods, the analysis spans the signal's entire duration, lacking the ability to provide temporal information about the timings of these patterns within the signal. In contrast, time-frequency methods \cite{Attoui2, tf12, tf16, tf22, tf27, tf32, tf35, tf43, aat22} transform the signal into energy-time-frequency representations where the signal's energy is mapped across both time and frequency. This, in turn, allows the identification of time-varying spectral characteristics within the signal. Common time-frequency transforms include short-time Fourier transform (STFT), Hilbert-Huang transform (HHT), and wavelet transform (WT). Despite their effectiveness, especially when handling nonstationary and nonlinear signals, time-frequency methods involve higher computational complexity--- in terms of online processing time and memory usage--- compared to pure spectral analysis. These factors directly influence the reliability of the monitoring process and associated financial expenses. Specifically, an increase in memory demand increases financial costs, while lengthy delays in condition prediction may fail to prevent costly catastrophic failures. In the context of VBCM of rolling bearings, vibrations of damaged bearings are manifested as modulations in the amplitude of the generated vibration signal \cite{env14}. This makes envelope-based analysis \cite{env1, env4, env5, env13, env14} an effective approach to facilitate efficient condition monitoring. In envelope-based analysis, the Hilbert transform (HT) is commonly employed to obtain instantaneous amplitude (IA) ``envelope," instantaneous phase (IP), and instantaneous frequency (IF). Accordingly, various feature extraction techniques can be used to extract relevant features from the obtained instantaneous information. \par

To this end, this paper introduces a low-complexity method for VBCM of rolling bearings based on joint analysis of IA and IF information of the generated vibration signal. Specifically, the proposed method employs the HT to obtain instantaneous information and then jointly analyze IA and IF information, facilitating new joint instantaneous amplitude-frequency representations of the vibration signal. Accordingly, six new fault-sensitive features are engineered from these representations. Besides the small number of extracted features, the proposed method uses very short durations of the generated vibration signal for condition monitoring, thereby relaxing memory requirements and reducing monitoring delay. The main contributions of the paper include:
\begin{itemize}
     \item Introducing a new low-complexity method for VBCM of rolling bearings based on envelope analysis.
     \item In the proposed method, the instantaneous amplitude ``envelope" and the instantaneous frequency of the vibration signal are jointly analyzed to facilitate three novel envelope representations: instantaneous amplitude-frequency mapping (IAFM), instantaneous amplitude-frequency correlation (IAFC), and instantaneous energy-frequency distribution (IEFD).
     \item Maintaining temporal information, the introduced representations effectively capture energy-frequency variations that are unique to the condition of the bearing, thereby enabling the extraction of discriminative features with high sensitivity to variations in operational conditions.
     \item Accordingly, six new highly discriminative features are extracted from these representations. 
     \item The extracted features are engineered to characterize shapes of the proposed instantaneous representations, thereby capturing instantaneous energy-frequency dynamics of the signal's envelope.
     \item The proposed method facilitates a low-complexity VBCM since it utilizes input vibration segments of very short durations (0.1 seconds) and produces six features only. Thus, relaxing memory requirements and reducing monitoring delays, which, in turn, helps reduce memory costs and prevent costly catastrophic failures.
\end{itemize}

The remainder of the paper is structured as follows: Related Work is presented in the next section. Section III introduces the proposed method. Section IV addresses performance evaluation in terms of the used dataset and experimental setup. Section V introduces and discusses the obtained results. The paper is finally concluded in Section VI.

\section{Related Work}
The HT is commonly employed in VBCM applications to obtain instantaneous information (IA, IP, and IF) of generated vibration patterns and, consequently, extract distinctive fault-related features  \cite{env1, env4, env5, env13, ch7_1, ch7_2, ch7_3, ch7_4, ch7_5, ch7_6, ch7_7, ch7_8, ch7_9, ch7_10, ch7_11, ch7_12, ch7_13, ch7_14, ch7_15, ch7_16, ch7_17, ch7_18, ch7_19, ch7_20, tf22, tf32, tf35}. The existing HT-based approaches for feature extraction can be generally grouped into three broad categories: Envelope-based spectral analysis \cite{env4, ch7_2, ch7_6, ch7_10, ch7_14, ch7_15, ch7_16, ch7_17}, signal decomposition and envelope reconstruction \cite{ch7_1, env5, ch7_3, ch7_5, ch7_8, ch7_9, ch7_11, ch7_12}, and time-frequency analysis \cite{env13, ch7_4, ch7_18, tf22, tf32, tf35, ch7_7, ch7_10, ch7_13, ch7_19, ch7_20}. Envelope-based spectral analysis involves analyzing the spectrum of the obtained envelope (IA) to extract fault-related features. While pure frequency analysis of the signal envelope is effective under steady-state conditions, it falls short in capturing the complexity of bearing vibrations under time-varying speed and load conditions, which is essential for real-time applications. \par

Signal decomposition and envelope reconstruction approaches utilize signal decomposition techniques such as wavelet decomposition and adaptive mode decomposition (AMD) to decompose the time-domain envelope into elementary modes. A screening process is then conducted to identify fault-informative modes. Finally, the signal is reconstructed using the identified modes only, and the envelope spectrum is obtained accordingly for fault analysis. The main advantage of decomposition-based approaches is that they attempt to reconstruct a low-redundancy and highly fault-sensitive envelope through decomposition, screening, and reconstruction processes. Additionally, selecting fault-related modes and discarding the remaining modes reduces noise presence in the reconstructed signal, thereby improving the signal-to-noise ratio (SNR)  of the obtained envelope compared to the original signal. However, such approaches involve high computational burdens due to decomposition, screening, and reconstruction processes. Further, proper mode screening criteria should be applied to avoid losing useful information. \par

Time-frequency analysis is particularly useful in analyzing nonstationary vibration patterns whose spectral properties change over time. In the context of HT-based VBCM, HHT and STFT are commonly adapted to perform time-frequency analysis of the signal envelope. Unlike envelope spectral analysis, which utilizes the Fourier transform (FT) to analyze the spectral contents of the envelope over its entire duration, STFT-based approaches analyze the spectral contents of the envelope over finite short-duration time windows, thereby preserving temporal information. A major limitation in STFT is its fixed segment `` window" length, which results in a uniform resolution analysis of the envelope, leading to an inherent compromise between time and frequency resolutions. HHT employs an alternative approach for time-frequency analysis through two steps: signal decomposition and Hilbert spectral analysis (HSA). In contrast to STFT, which relies on fixed-time durations and sinusoidal kernel functions for envelope analysis, HHT utilizes an adaptive approach to analyze the vibration signal, making it highly responsive to variations in vibration patterns. Specifically, HHT uses AMD techniques to decompose the signal into a set of simpler modes known as intrinsic mode functions (IMFs), representing different frequency components of the signal. After the decomposition process, HSA is conducted, where the HT is applied to each mode to obtain its IA and IF information, providing a detailed energy-time-frequency representation of the original signal commonly known as the Hilbert spectrum. The adaptive mechanism of HHT and the use of HSA makes it very effective in analyzing complex vibration patterns. The performance of the HHT, however, is highly dependent on the reliability of the decomposition process and its parameters, such as the stopping criterion for the decomposition process and the interpolation method for envelope estimation. Moreover, HTT involves high computational complexity due to its adaptive decomposition mechanisms, which makes it unsuitable for real-time VBCM applications. \par

This paper introduces a low-complexity method for VBCM of rolling bearings based on envelope analysis. The proposed method jointly utilizes the IA and IF of the vibration signal to facilitate three novel envelope representations that maintain temporal information to capture energy-frequency variations in the signal envelope effectively. The following section introduces the theoretical background and the details of the proposed method.

\section{Joint Instantaneous Time-Frequency Analysis of Vibration Signals}
 
\subsection{The Hilbert Transform}
The Hilbert transform (HT) is a fundamental operator in signal processing since it provides an efficient way to obtain analytic signal representations of real-valued signals. An analytic signal is a complex-valued representation of a signal that describes its amplitude and phase characteristics. Given signal $x(t)$, its HT is defined as:
\begin{equation}
    H\{x(t)\} = \frac{1}{\pi} \int_{-\infty}^{\infty} \frac{x(\tau)}{t-\tau} d\tau
\end{equation}
The transform essentially modifies the phase of each frequency component of the signal by $\pm 90^\circ$. Accordingly, the analytical signal $x_a(t)$ of $x(t)$ is formed by augmenting the signal with its HT $H\{x(t)\}$ as the imaginary part:
\begin{equation}
    x_a(t) = x(t)+jH\{x(t)\}
\end{equation}
Thus, the instantaneous amplitude (IA) or envelope, $A(t)$  of the signal is given by:\newline
\begin{equation}
    A(t) = |x_a(t)| = \sqrt{x(t)^2 + H\{x(t)\}^2},
\end{equation}
 and  the instantaneous phase (IP), $\theta(t)$ is expressed as:
\begin{equation}
    \theta(t) = \arctan\left(\frac{H\{x(t)\}}{x(t)}\right)
\end{equation}
Hence, the instantaneous frequency (IF), $F(t)$ can be calculated by taking the derivative of  the instantaneous phase $\theta(t)$:
\begin{equation}
    F(t) = \frac{1}{2\pi} \frac{d\theta(t)}{dt}
\end{equation}\

\subsection{Proposed Joint Instantaneous Time-Frequency Analysis}
\begin{figure}[!htbp]
\centerline{\includegraphics[width=0.5\textwidth]{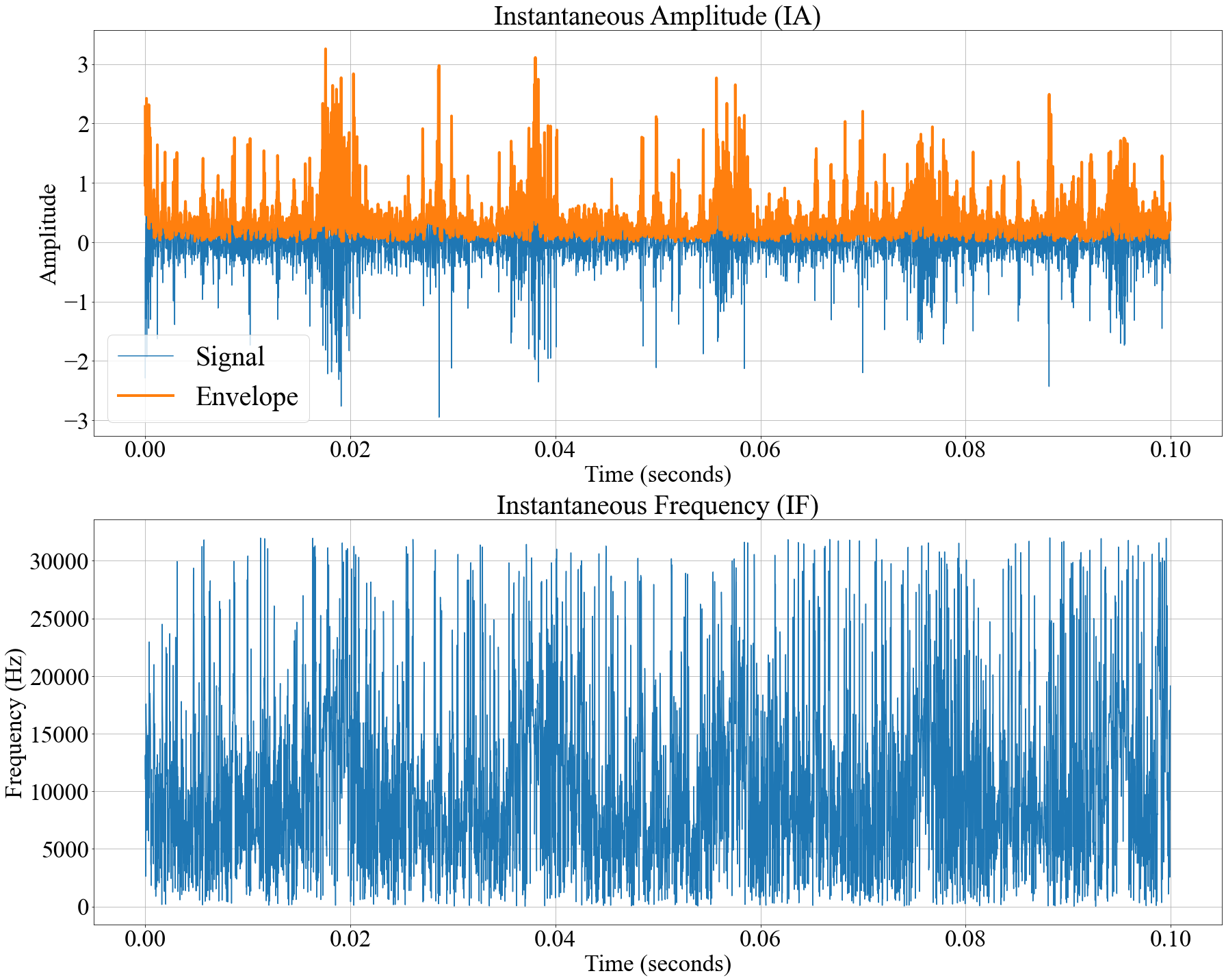}}
\caption{Instantaneous amplitude (IA) and Instantaneous frequency (IF) of a rolling bearing vibration signal, $v_h(t)$.}
\label{vib_ia_if}
\end{figure}

Fig. \ref{vib_ia_if} shows IA and IF information, obtained through the HT, of a real-world vibration signal of a rolling bearing, denoted as $v_h(t)$. The signal is extracted from the Paderborn University (PU) bearing dataset \cite{paderborn}. $v_h(t)$ represents a healthy vibration pattern and contains $6,400$ datapoints that were acquired at a sampling rate of $64,000$ Hz; thus, the signal spans a $0.1$ seconds of time duration.
As shown, the IA and IF information show how the signal's amplitude and frequency evolve over time, respectively. Hence, they separately describe the signal's time-energy and time-frequency characteristics, which are of particular significance since faults often manifest as changes in the vibration signal's energy and/or frequency content. Accordingly, in this paper, IA and IF of the vibration signal are jointly utilized to conduct an efficient instantaneous amplitude-frequency analysis of the signal envelope. The joint instantaneous time-frequency analysis is facilitated through three novel envelope representations: Instantaneous amplitude-frequency mapping (IAFM), instantaneous amplitude-frequency correlation (IAFC), and instantaneous energy-frequency distribution (IEFD). Consequently, six highly discriminative features are extracted from these representations,  characterizing the shapes of the proposed representations and capturing instantaneous energy-frequency dynamics of the signal’s envelope.

\subsubsection{Instantaneous Amplitude-Frequency Mapping}
\begin{figure}[!htbp]
\centerline{\includegraphics[width=0.5\textwidth]{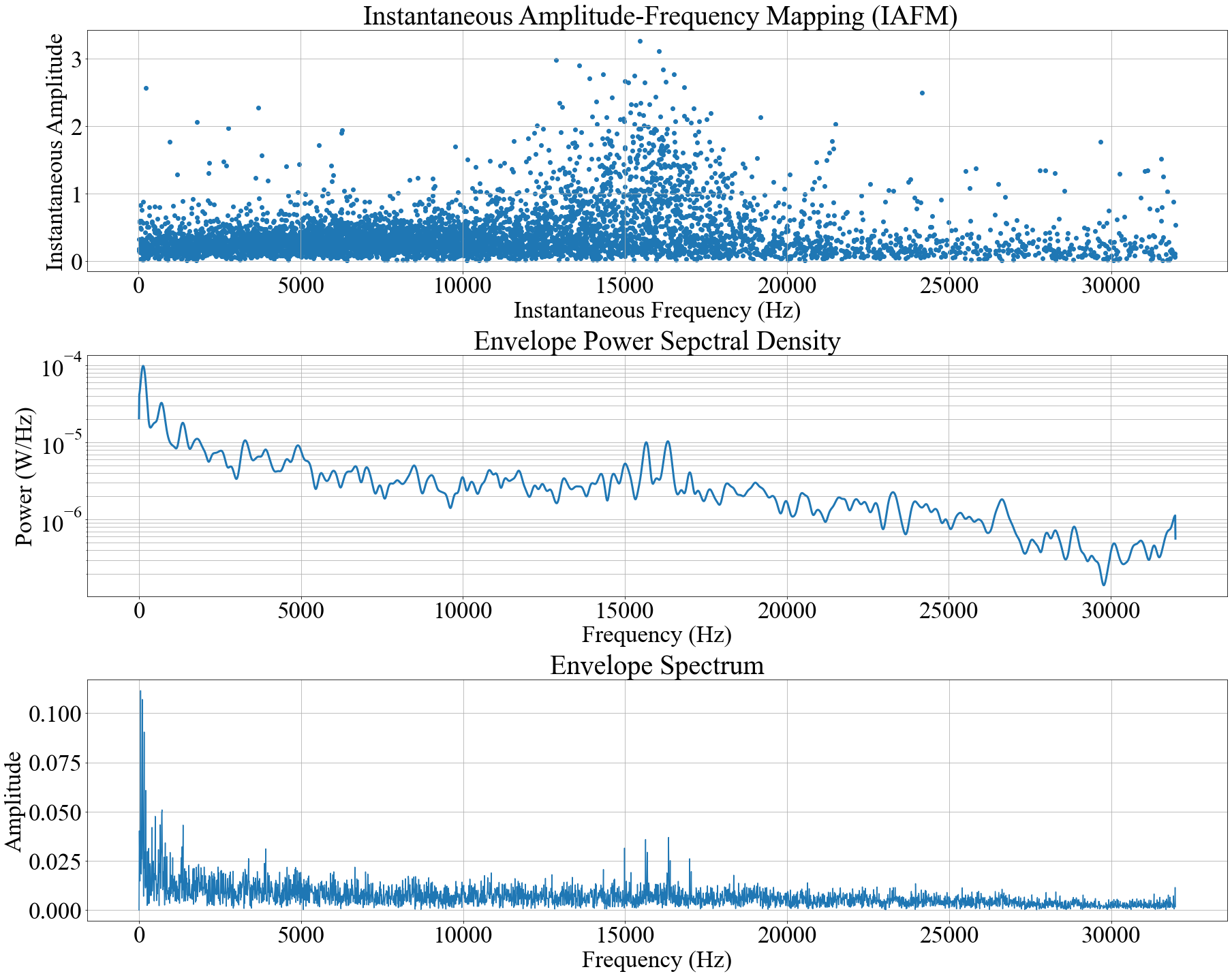}}
\caption{Instantaneous amplitude-frequency mapping (IAFM), Envelope PSD, and envelope spectrum of the healthy vibration pattern, $v_h(t)$.}
\label{iafm_psd_fft}
\end{figure}

Given IA and IF information, the instantaneous amplitude-frequency mapping (IAFM) is reconstructed by mapping IA and IF information together where the $x$-axis represents the IF, and the $y$-axis represents the IA as depicted in Fig. \ref{iafm_psd_fft}. The figure also displays the PSD and frequency spectrum of the envelope.
A comparison among the three spectral representations reveals that the proposed IAFM effectively captures the spectral shape of the envelope. In PSD and frequency spectrum, each $x-y$ pair represents a unique frequency-energy value across the spectral representation, which is helpful in precisely identifying dominant frequency components. In contrast, as instantaneous amplitude-frequency mapping, the IAFM shows how signal amplitude ``energy" values are allocated through various frequencies over the signal's time duration, thereby providing an energy-frequency density representation of the envelope with high energy resolution. This makes IAFM particularly useful in characterizing energy-frequency behaviors of healthy and faulty vibration patterns through the spread and concentration of energy ``amplitude" values across various frequencies. 
Accordingly, the following features are extracted from IAFM that capture and characterize its shape:
\begin{itemize}
    \item Spectral Centroid (SC): The SC represents the center of gravity of the IAFM and quantifies the average frequency at which the energy is concentrated. It describes the spectral position of dominant oscillations in the generated vibration, thereby identifying various vibration patterns. Given that IA and IF information are represented in the discrete forms $A[n]$ and $F[n]$, respectively, SC is calculated through the following formula:
    \begin{equation}
         CS = \frac{\sum_{n=0}^{N-1} F[n] \cdot A[n]}{\sum_{n=0}^{N-1} A[n]} \;\textit{Hz}
    \end{equation}
    where $N$ is the number of total datapoints in $A[n]$ and $F[n]$.
    \item Spectral Spread (SS): It describes the deviation of instantaneous amplitude-frequency points in the IAFM  with respect to the SC. Hence, it provides a measure to assess the variance of the IAFM. SS is obtained by calculating the standard deviation of IAFM with respect to SC:
    \begin{equation}
         SS = \sqrt{\frac{\sum_{n=0}^{N-1} (F[n]-CS)^2 \cdot A[n]}{\sum_{n=0}^{N-1} A[n]}} \;\textit{Hz}
    \end{equation}
    \item Coefficient of Variation (CoV): This metric provides insights into the dispersion of the envelope's energy across its frequency components. CoV is expressed as the ratio of the SC to the CC:
    \begin{equation}
         CoV = \frac{SS}{SC}\times100 \; (\%)
    \end{equation}
    CoV quantifies the dispersion of the IAFM with respect to SC, indicating how spread out the energy is in relation to the mean frequency. A higher CoV implies a wider distribution of the envelope's energy across its instantaneous spectrum, suggesting a more dispersed frequency content. Conversely, a lower CoV indicates a more concentrated distribution of frequencies around the SC, implying that the envelope's energy is more narrowly focused around a central frequency.
\end{itemize}

\subsubsection{Instantaneous Amplitude-Frequency Correlation}
The second proposed representation is the instantaneous amplitude-frequency correlation(IAFC), which represents the cross-correlation of IA and IF:
\begin{equation}
    \text{IAFC} = R_{AF}[k]=\sum_{n=0}^{N-1}A[n]\cdot F[n-k], \; k=0, \pm1,\dots,\pm N-1
\end{equation}
\begin{figure}[!htbp]
\centerline{\includegraphics[width=0.5\textwidth]{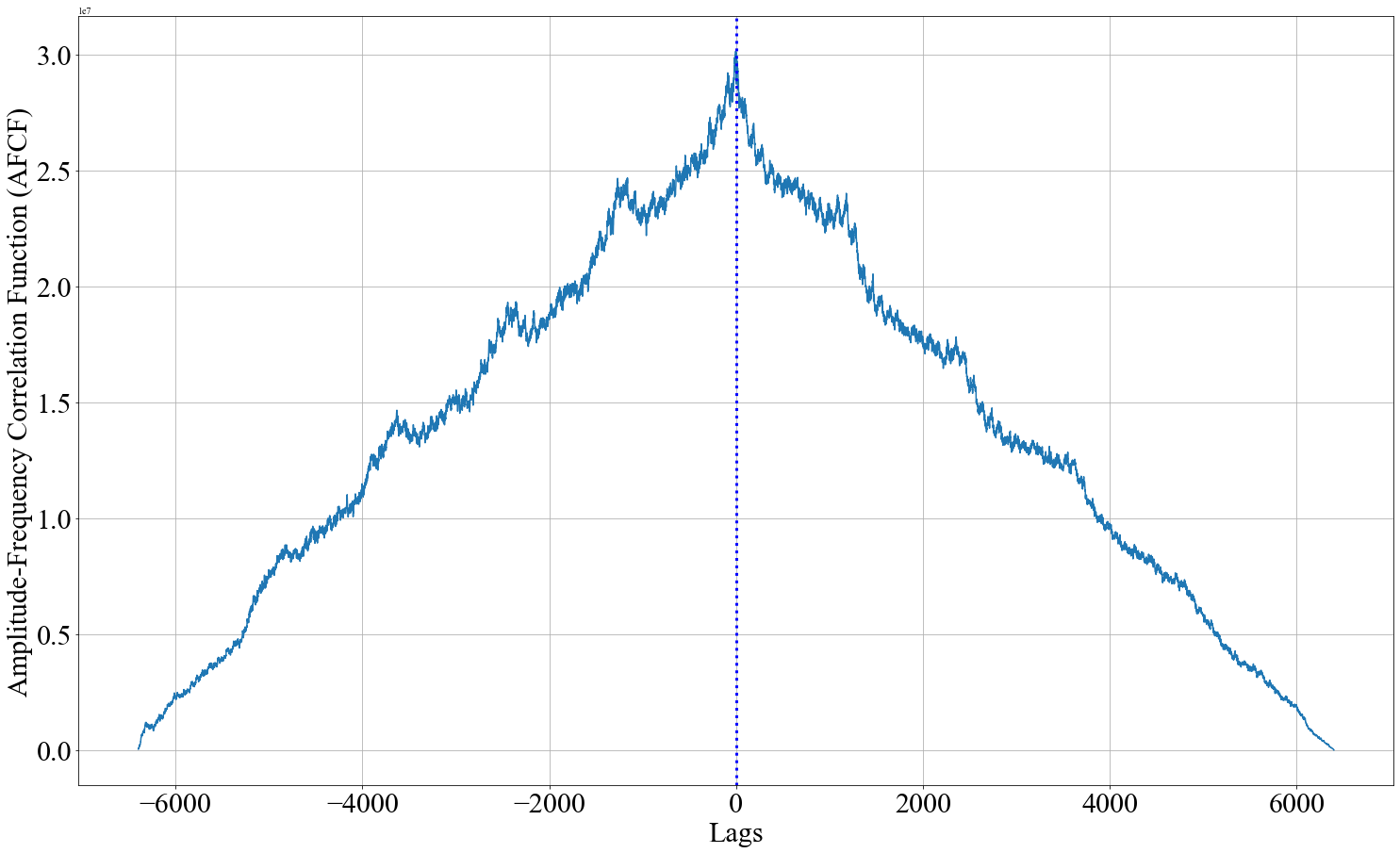}}
\caption{IAFC of the healthy vibration pattern, $v_h(t)$}
\label{afcf}
\end{figure}
As instantaneous quantities, IA and IF maintain the same temporal dependencies among their values. Accordingly, the IAFC would capture the mutual dynamics between IA and IF that are unique to the system's health state since changes in operational conditions often manifest as variations in amplitude and frequency. Thus, the IAFC can be interpreted as an envelope representation that describes the relationship between the envelope's energy and frequency variations over time. Since  IA and IF are both positive quantities, the IAFC would possess common characteristics as well as condition-specific features. Specifically, given the IAFC of the healthy vibration pattern, $v_h(t)$ depicted in Fig.\ref{afcf}, the following observations are made:
\begin{enumerate}
    \item The IAFC is a positive function since IA and IF are positive quantities.
    \item The IAFC has an increasing trend until it reaches its peak, after which it shows a decreasing trend.
    \item The peak, representing the maximum cross-correlation,  typically occurs at full overlap between IA and IF ``zero lag". Otherwise, it reflects very high excitements in either amplitude, frequency, or both, which would indicate abnormal behavior.
\end{enumerate}
Accordingly, the following features are extracted from the IAFC to quantify the aforementioned characteristics, thereby capturing various operation conditions:
\begin{itemize}
    \item Correlation Peak (CP): It represents the maximum value of IAFC, quantifying the maximum cross-correlation. CP is expressed mathematically as follows:
    \begin{equation}
       CP = max\left \{ R_{AF}[k] \right \}, \;k=0, \pm1,\dots,\pm N-1
    \end{equation}
    \item Peak Lag (PL): PL is the lag at which the $R_{AF}[k]$ has its maximum value. In other words, it is the lag value corresponding to the CP. 
\end{itemize}
The values of CP and PL characterize the maximum cross-correlation in the IAFC, which would be a discriminative indicator between healthy and faulty conditions, as explained above.

\subsubsection{Instantaneous Energy-Frequency Distribution}
\begin{figure}[!htbp]
\centerline{\includegraphics[width=0.5\textwidth]{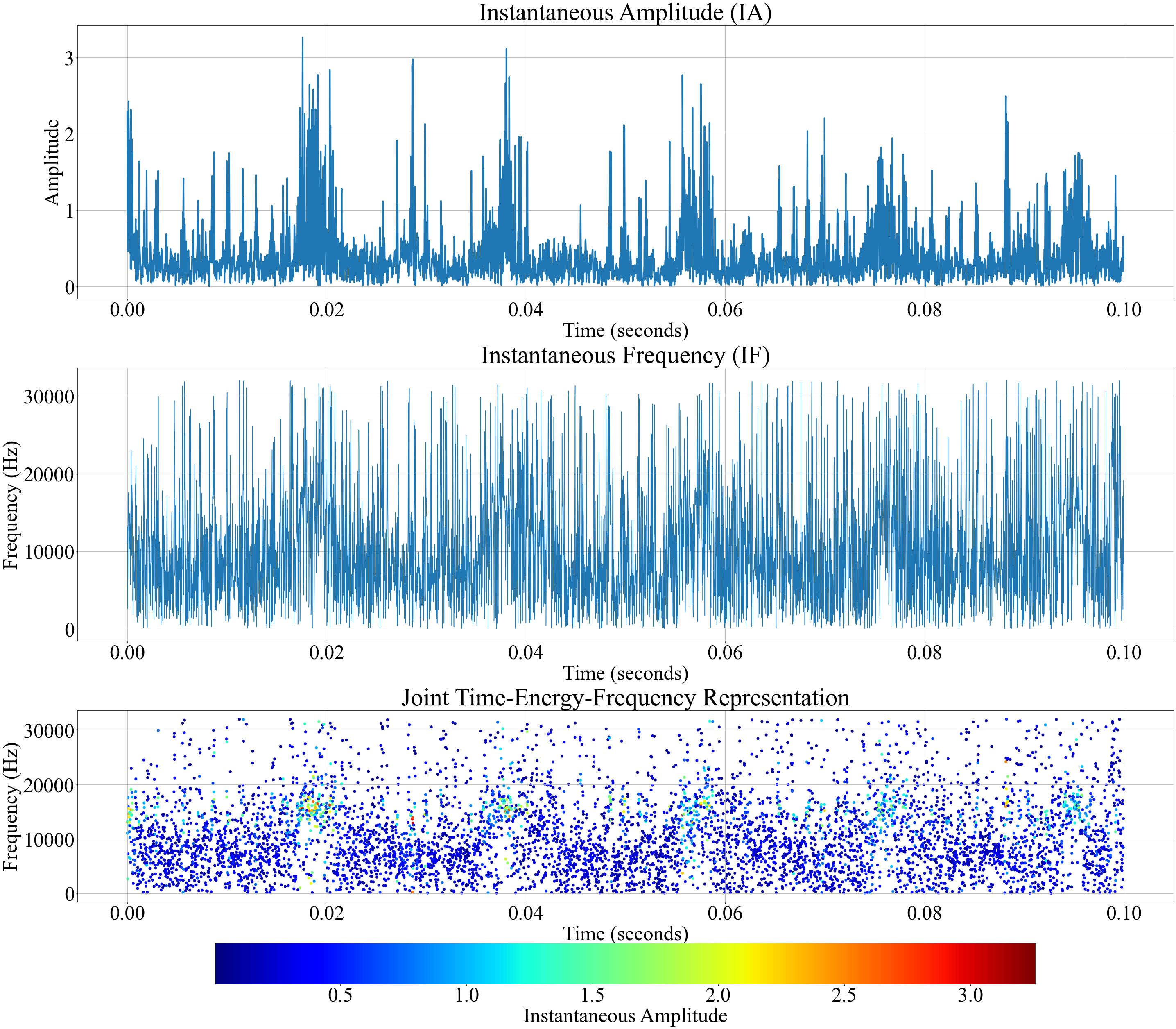}}
\caption{Joint time-energy-frequency representation of the healthy vibration pattern, $v_h(t)$}
\label{heatmap}
\end{figure}

\begin{figure}[!htbp]
\centerline{\includegraphics[width=0.5\textwidth]{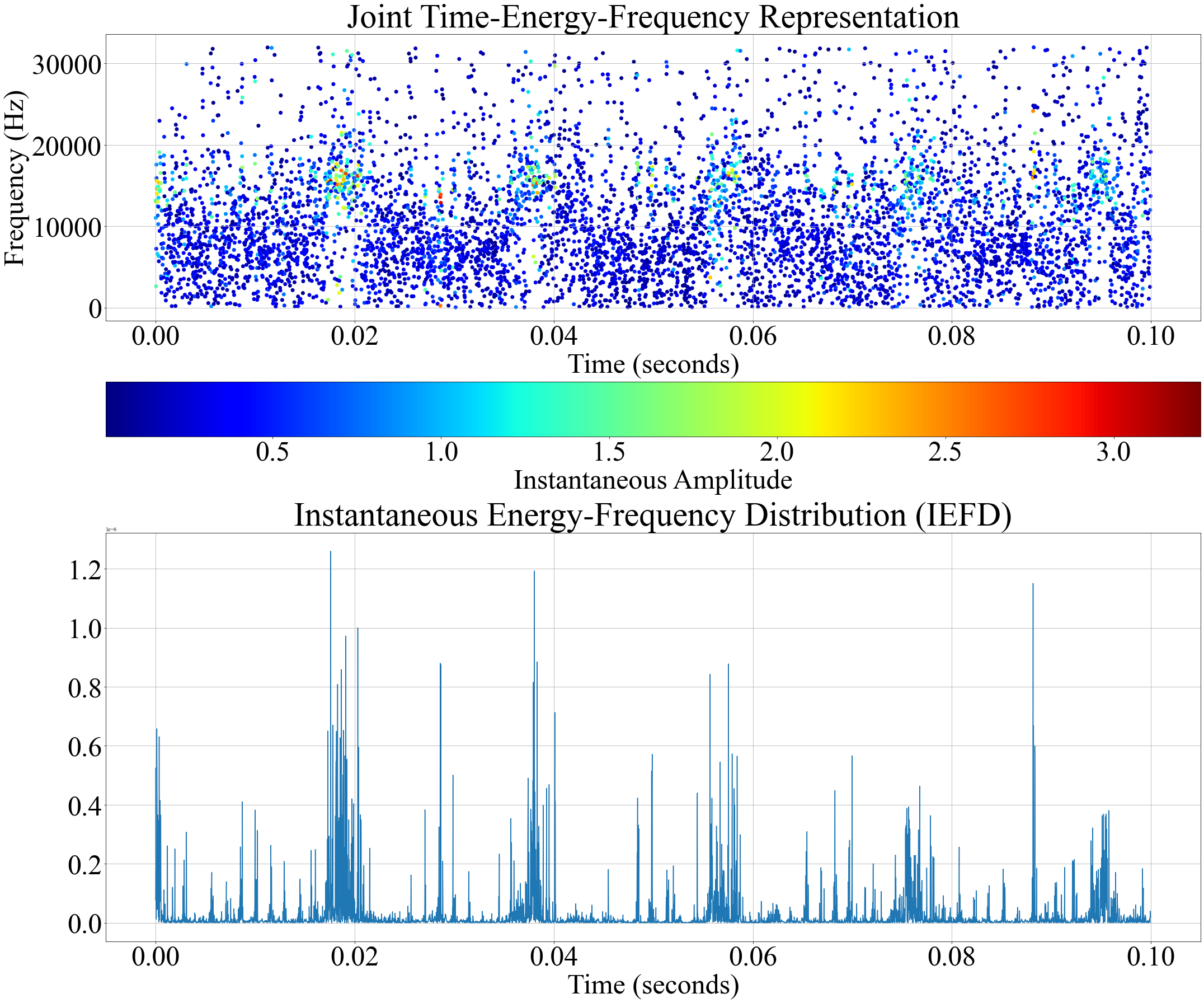}}
\caption{IEFD along with the joint time-energy-frequency visualization of the healthy vibration pattern, $v_h(t)$.}
\label{iefd}
\end{figure}

As previously mentioned, IA and IF information independently show how the envelope's energy and spectral content evolve over time. Specifically, while IA shows how the energy is distributed over the time period, the IF shows how the frequency changes over that time period. Accordingly, a joint time-energy-frequency representation of the signal can be visualized by mapping the intensity of each time-frequency pair in the IF to its corresponding value of the IA, as demonstrated in Fig. \ref{heatmap} for the healthy vibration pattern, $v_h(t)$.
This visual heatmap is an effective tool for visualizing the instantaneous components since it clearly represents the variations in the envelope's energy and frequency over time. To facilitate this visualization efficiently, we introduce the third proposed representation, Instantaneous Energy-Frequency Distribution (IEFD), which is the product of normalized instantaneous energy ($IE_{norm})$ and normalized IF ($IF_{norm}$). The $IE_{norm}$ is obtained by squaring and normalizing the instantaneous amplitude $A[n]$:
\begin{equation}
    IE_{norm}[n] = \frac{A[n]^2}{\sum_{n=0}^{N-1}A[n]^2}
\end{equation}
The squaring and normalization processes emphasize larger values of IA  than smaller ones and transform the instantaneous amplitude values into values that represent their proportion to the envelope's total energy over the given time period. Thus, $IE_{norm}$ is more sensitive to points in time where the energy is significantly higher, as these points will contribute disproportionately to the sum. The $IF_{norm}$ is obtained by normalizing the instantaneous frequency $F[n]$:
\begin{equation}
    IF_{norm}[n] = \frac{F[n]}{\sum_{n=0}^{N-1}F[n]}
\end{equation}
Normalizing IF makes the representation more sensitive to changes in the frequency content of the envelope. Specifically, the normalization process provides a measure of the relative contribution of each instant's frequency to the overall envelope’s spectrum, thereby amplifying the impact of large frequency values in fault analysis. Accordingly, the IEFD is expressed as follows: 
\begin{equation} 
\begin{split}
 IEFD[n] & = IE_{norm}[n] \times IF_{norm}[n] \\
 & = \frac{A[n]^2}{\sum_{n=0}^{N-1}A[n]^2} \times \frac{F[n]}{\sum_{n=0}^{N-1}F[n]}
\end{split}
\end{equation}
By combining the normalized instantaneous energy and frequency information, the IEFD provides a comprehensive representation that captures crucial dynamics of the generated vibration. As a result, it allows localized analysis in the time-energy-frequency domain, thereby enhancing the detection of changes or shifts in operational behavior that can be used as indications of faults. Fig. \ref{iefd} shows and compares the IEFD to the joint time-energy-frequency visualization of the healthy vibration pattern, $v_h(t)$.
The comparison shows that the IEFD effectively captures the time-varying energy-frequency characteristics of the signal's envelope. These characteristics are highly influenced by changes in the operational conditions since the presence of a fault is often characterized by a high level of regularity in the generated vibration due to the emergence of highly energetic frequency components associated with the fault. Conversely, normal operation conditions would generate more complex and highly irregular vibration patterns. To quantify these characteristics and extract a feature that is sensitive to such changes in operational conditions, we introduce the mean-to-entropy ratio (MER) of the IEFD. The MER represents the ratio between the mean and Shannon entropy of IEFD; it is expressed mathematically as follows:
\begin{equation}
    MER = \frac{\frac{\sum_{n=0}^{N-1}IEFD[n]}{N}}{-\sum_{i=0}^{M-1} x_i \log_2 P(x_i)}
\end{equation}
where $i = 0, 1,\cdots M-1$, $M$ is the number of unique values in the IEFD, and $P(x_i)$ is the probability of each unique value $x_i$. The mean value of the IEFD quantifies its average concentration in terms of joint energy-frequency content across the observed time period, while entropy quantifies irregularity within the IEFD. \par

To this end, instantaneous amplitude and frequency information of vibration signals are jointly utilized to facilitate three novel envelope representations. Accordingly, six highly discriminative features are extracted from these representations as summarised in Table \ref{table_1}.

\subsubsection{Complexity Analysis}
Algorithm \ref{alg:alg1} shows the pseudo-code and computation steps of the proposed method.
\begin{table}[!b]
  \caption{Proposed envelope representations and extracted features}
  \label{table_1}
  \includegraphics[width=0.5\textwidth]{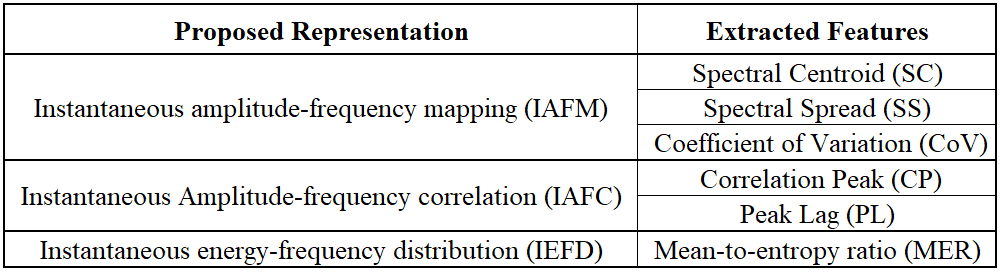}
\end{table}
\begin{algorithm}
	\caption{Proposed Joint Instantaneous Time-Frequency Analysis} 
	\begin{algorithmic}
		\State \textit{Input:} vibration signal $x[n]$ of length $N$ data points
		\State \textit{Parameters:} $f_s$ = sampling frequency (samples/second).
		\State \textit{Output:} $\mathcal{F}[s], s= 0,...,S-1:$ features vector of size $S=6$
		\State \textbf{Start:}
	    \State Compute Hilbert Transform (HT) of $x[n]$: $H\left \{ x[n] \right \} =  HT(x[n], fs)$.
		\State Compute instantaneous amplitude $A[n]$ and instantaneous frequency $F[n],\;n = 0,1,\cdots N$. \newline
		\% 1- Instantaneous amplitude-frequency mapping (IAFM):
            \State Calculate SS, $\mathcal{F}[0] \gets SS$
            \State Calculate SC, $\mathcal{F}[1] \gets SC$
            \State Calculate CoV, $\mathcal{F}[2] \gets CoV$ \newline
            \% 2- instantaneous Amplitude-frequency correlation function (IAFC):
		\State Calculate CP, $\mathcal{F}[3] \gets CP$ 
		\State Calculate PL, $\mathcal{F}[4] \gets PL$  \newline
            \% 3- Instantaneous energy-frequency distribution (IEFD):
            \State Calculate MER, $\mathcal{F}[5] \gets MER$ 
		\State \textbf{End}
	\end{algorithmic}
	\label{alg:alg1}
\end{algorithm}
Accordingly, the complexity of the proposed method can be analyzed as follows:
\begin{itemize}
    \item Complexity of the HT is $O(N\log{}N)$ since it is commonly implemented using the fast Fourier transform (FFT) \cite{ch7_21, ch7_22}, which is the case in both MATLAB and Python SciPy implementations of the HT.
    \item Complexity of the cross-correlation function IAFC is $O(N\log{}N)$ considering FFT-based cross-correlation.
    \item Complexity of the other computations are in the order of $O(N)$.
\end{itemize}
Thus, the most computationally intensive steps are those involving the HT and the IAFC, each with a complexity of $O(N\log{}N)$. Therefore, the overall complexity of the proposed method is dominated by these steps, leading to an overall computational complexity of $O(N\log{}N)$.

\section{Performance  Evaluation}
\subsection{Experimental Dataset}
The performance of the proposed method is evaluated on the Paderborn University (PU) bearing dataset \cite{paderborn}. In contrast to other datasets, the PU dataset has real bearing damages with combined defects. The measurements are conducted using a 425 $W$ Permanent Magnet Synchronous Motor (PMSM). The dataset used in this paper is based on measurements conducted at n = 1,500 rpm with a load torque of M = 0.7 Nm and a radial force on the bearing of F = 1,000 N. Vibration signals are recorded with a sampling rate of 64,000 Hz by measuring the acceleration of the bearing housing at the adapter at the top end of the rolling bearing module. Regarding bearing defects, the PU dataset includes artificially generated and accelerated-lifetime defects. In this paper, only accelerated-lifetime defects are used. Accordingly, the dataset has four classes: one healthy class and three faulty classes according to fault type, as shown in Table \ref{table_2}.
\begin{table*}
\caption{Experimental dataset}
\begin{tabular}{|c|c|c|c|c|c|c|c|}
\hline
\multicolumn{1}{|c|}{\multirow{5}[1]{*}{\textbf{ PU dataset}}} & \textbf{Class} & \textbf{Health condition} & \textbf{Fault type} & \multicolumn{4}{c|}{\textbf{Motor speed (rpm)}} \\
\cline{2-8}      & 1     & Healthy & \multicolumn{1}{c|}{NA} & \multicolumn{4}{c|}{\multirow{4}[1]{*}{1500}} \\
\cline{2-4}     & 2    & Combined IR and OR faults  & Multiple damages & \multicolumn{4}{c|}{} \\
\cline{2-4}     & 3    & IR faults & Single, repetitive, and multiple damages & \multicolumn{4}{c|}{} \\
\cline{2-4}     & 4    & OR faults & Single and repetitive damages & \multicolumn{4}{c|}{} \\
\hline
\end{tabular}%
\centering
\label{table_2}
\end{table*}
\subsection{Experimental Setup}
In the preprocessing stage, vibration signals of the dataset are segmented into non-overlapping segments of $N = 6,400$ samples. Accordingly,  the resultant segment duration is $0.1$ seconds, given that the vibration signals are acquired at $64,000$ Hz. The segment duration of $0.1$ seconds is precise enough to facilitate real-time monitoring with moderate computational requirements. The segmentation process results in a dataset of $16,005$ vibration segments in total. After extracting the features, the resulting dataset is divided into two parts: $11,203$ samples for training (70\%) and $4,802$ samples for testing (30\%). A random forest (RF) classifier is then trained on the training dataset. The selection of RF is driven by its better performance compared to other classifiers, as reported in \cite{aat22}. The performance of the proposed method is compared with two common time-frequency methods: the short-time-Fourier transform (STFT) of the signal envelope and The Hilbert-Huang transform (HHT) of the signal. In the STFT-based method, the envelope of the vibration signal is segmented into three segments using the Hamming window with $50\%$ overlap.  The FFT is then computed for each segment with the number of FFT points (NFFT) set equal to $4,096$. The proposed features are calculated accordingly considering that $A[n]$ represents aggregated STFT power coefficients and $F[n]$ represents frequency bins. In the HHT-based method, the obtained IA and IF information of the resulting intrinsic mode functions (IMFs) are aggregated, and the proposed features are calculated accordingly. The following metrics are used to evaluate the performance:
\begin{itemize}
    \item Prediction accuracy (\%) and ROC-AUC score to evaluate the reliability and effectiveness of the proposed method in condition monitoring.
    \item Online processing time and memory usage to assess computational complexity of the proposed method.  
\end{itemize}
Python programming language, along with \textit{SciPy} and \textit{emd} libraries, are used to build the models. The code is publicly available on the Github site of the Optimized Computing and Communications (OC$^2$) Laboratory\footnote{https://github.com/Western-OC2-Lab/Joint-Instantaneous-Amplitude-Frequency-Analysis-for-Vibration-Based-Condition-Monitoring}. The achieved results and related discussion are presented in the next section.

\section{Results and Discussion}
\begin{figure*}[!htbp]
\centerline{\includegraphics[width=0.7\textwidth]{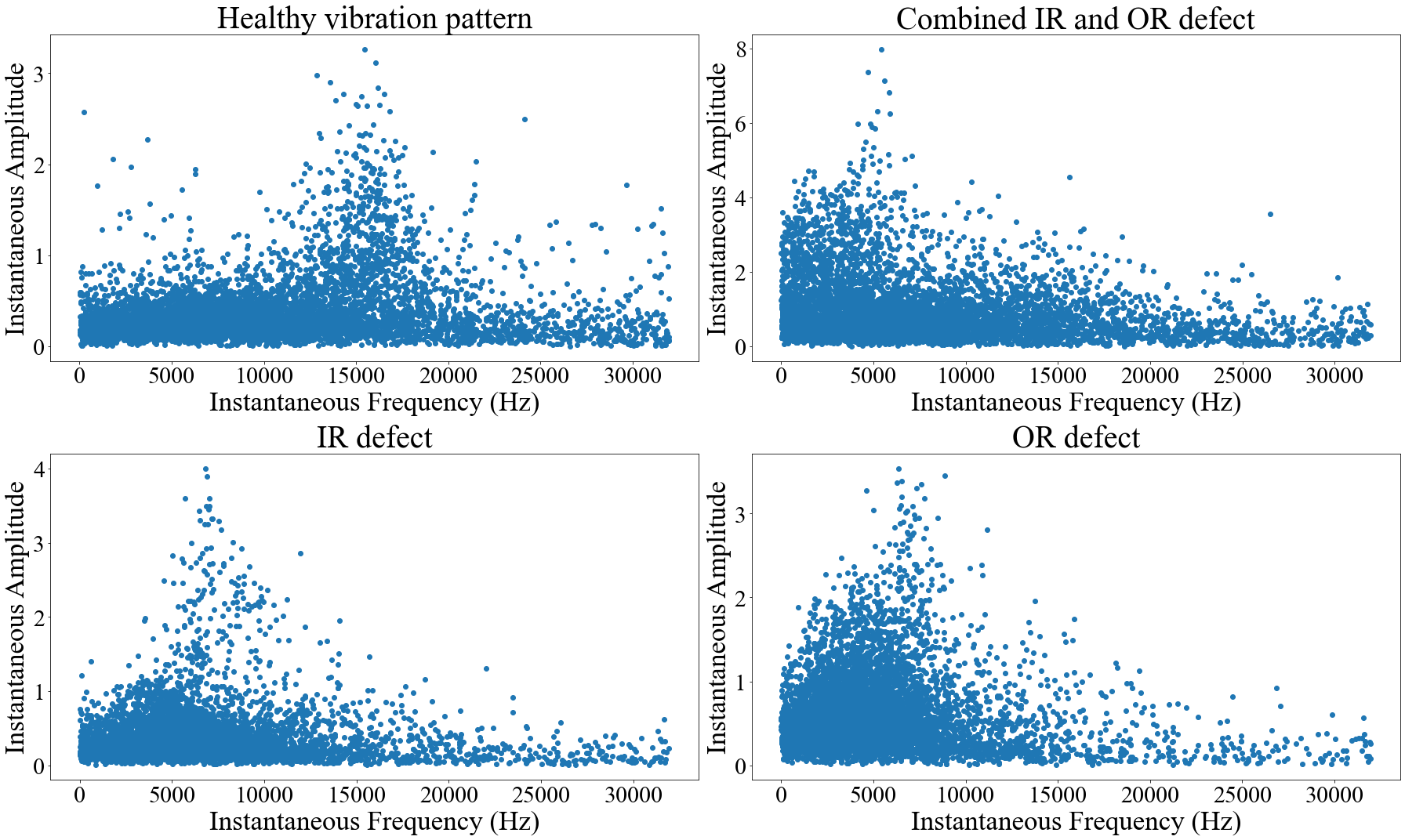}}
\caption{IAFMs of healthy, inner race (IR) defect, outer race (OR) defect, and combined IR and OR defect vibration patterns.}
\label{iafm_all}
\end{figure*}

\begin{figure}[!htbp]
\centerline{\includegraphics[width=0.5\textwidth]{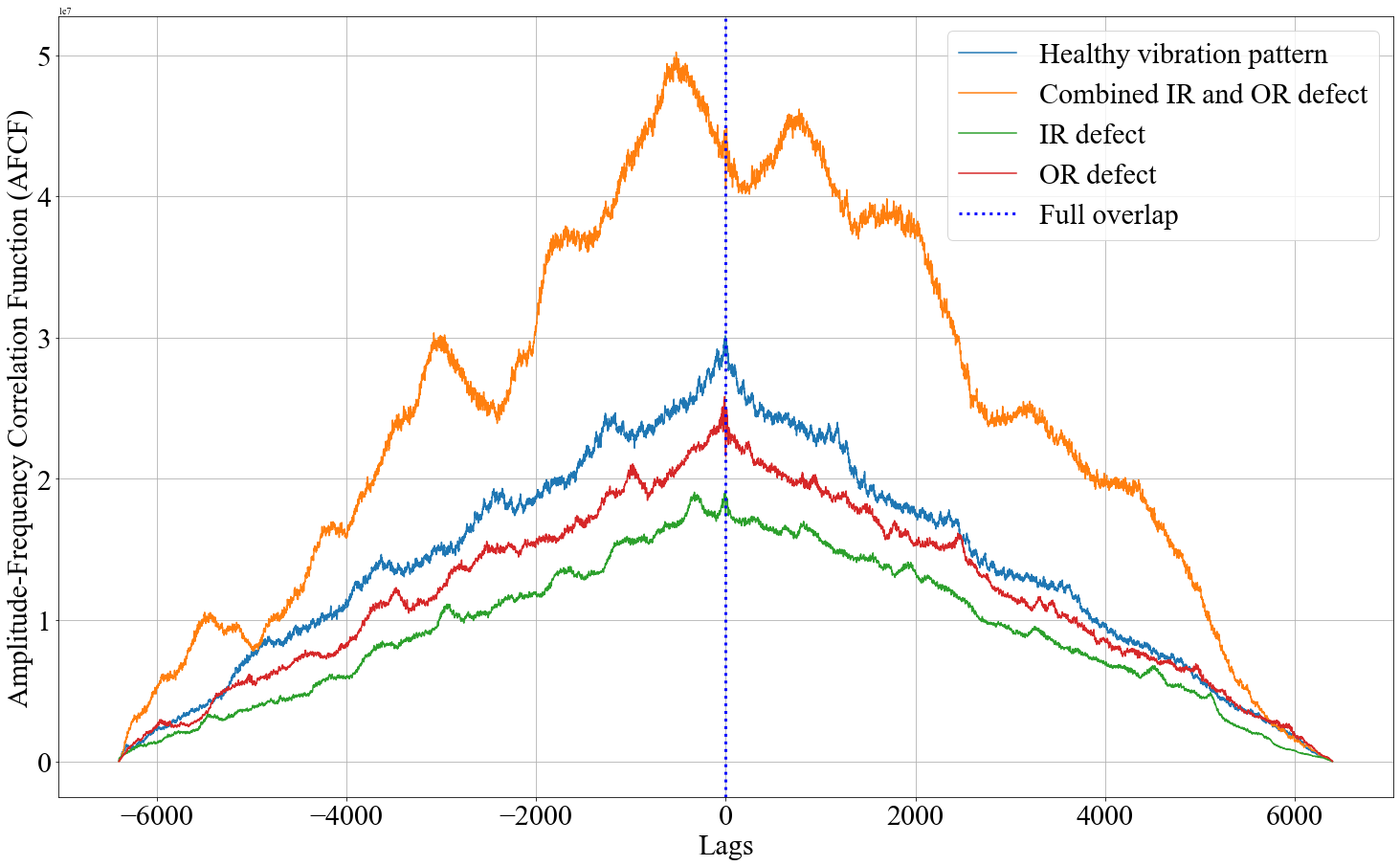}}
\caption{IAFCs of healthy, inner race (IR) defect, outer race (OR) defect, and combined IR and OR defect vibration patterns.}
\label{afcf_all}
\end{figure}

\begin{figure*}[!htbp]
\centerline{\includegraphics[width=0.7\textwidth]{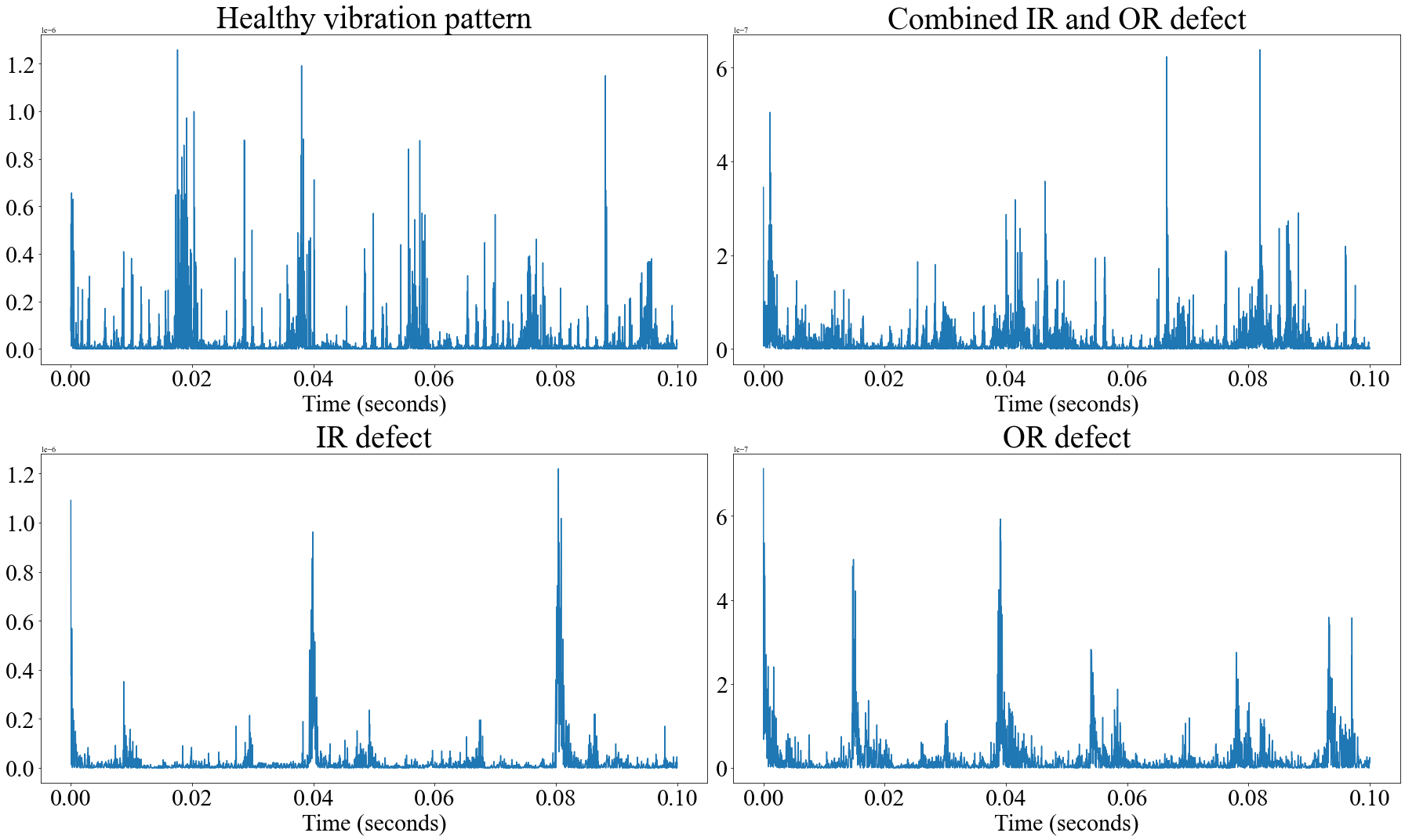}}
\caption{IEFDs of healthy, inner race (IR) defect, outer race (OR) defect, and combined IR and OR defect vibration patterns.}
\label{iefd_all}
\end{figure*}

Fig. \ref{iafm_all}, Fig. \ref{afcf_all}, and Fig. \ref{iefd_all} compare between IAFMs, IAFCs, and IEFDs, respectively, across healthy and faulty vibration patterns from the PU dataset. 
It is evident that the proposed envelope representations distinctly capture the unique variations between healthy and faulty bearing conditions by emphasizing the contrast in energy and frequency among various operational conditions. This, in turn, improves the discriminative characteristics of the extracted features which is crucial for facilitating effective condition monitoring. \par

Table \ref{table_3} shows the achieved performance results of the proposed method along with results of HHT-based features and STFT-based features. 
\begin{table*}[!htbp]
\caption{Performance comparison among the proposed method, the STFT-based method, and the HHT-based method.}
\begin{tabular}{c c c c c}
\hline
Method & Accuracy (\%) & ROC-AUC score & Online processing time (seconds) & Memory usage (MB) \\
\hline
Proposed & 99.60\% &  1.00 &  0.16 &  1.1  \\
STFT-based & 95.20\% &  0.997 &  0.16 & 0.9 \\
HHT-based & 91.46\% &  0.989 &  0.24 & 2.5\\
\end{tabular}
\centering  
\label{table_3}
\end{table*}
In terms of accuracy and ROC-AUC score, the proposed method achieved superior performance (accuracy $>99.6\%$, ROC-AUC score $=1.00$), reflecting its reliability and effectiveness in the detection and diagnosis of various fault types. Further, the results demonstrate the effectiveness of the proposed instantaneous amplitude-frequency representations in extracting highly fault-sensitive features compared to the other two methods. In contrast to the STFT-based method, which facilitates time-varying spectral representation of the signal envelope, the prospered method effectively utilizes instantaneous frequency, along with the envelope, to jointly facilitate instantaneous amplitude-frequency representations. As the results show, the proposed method outperforms the STFT-based method, indicating that the proposed presentations effectively capture more unique variations in energy and frequency between healthy and faulty bearings compared to the STFT-based representation. This could be explained by the limitation of fixed segment length in STFT, which results in an inherent compromise between time and frequency resolutions in the time-energy-frequency representation. Regarding HHT, although it utilizes instantaneous amplitude and frequency information to facilitate time-energy-frequency analysis, it results in lower accuracy compared to the other two methods. This gap in performance can be justified by impairment caused by the EMD process, such as mode mixing, end effects, and over-sifting. Regarding computational complexity, both the proposed method and the STFT-based methods demonstrate comparable and moderate computational requirements compared to the HHT-based method. This is expected since the iterative sifting process used in EMD to extract the IMFs is computationally intensive, especially for long vibration segments. \par 

Table \ref{table_4} presents a comparison, in terms of achieved prediction accuracy (\%),  with recent deep learning (DL)-based methods on the PU dataset. 
\begin{table*}
\caption{Performance comparison between the proposed method and other DL-based methods on the PU dataset.}
\begin{tabular}{c c c}
\hline
Method & Approach & Achieved accuracy (\%) \\
\hline
Proposed & Signal processing-based & 99.60\%  \\
\cite{comp1} & DL-based & 99.44\% \\
\cite{comp2} & DL-based  & 97.05\% \\
\cite{comp3} & DL-based & 96.51\% \\
\cite{comp4} & DL-based & 99.50\% \\
\cite{comp5} &  DL-based & 96.67\% \\
\cite{comp6} & DL-based & 97.68\% \\
\cite{dwangn} & DL-based & 99.80\% \\
\cite{akq} &  DL-based & 96.24\% \\
\cite{lhi} & DL-based & 99.70\% \\
\end{tabular}
\centering  
\label{table_4}
\end{table*}
The results demonstrate the effectiveness of the proposed method in achieving excellent performance compared to DL-based methods. Moreover, the proposed method is more computationally efficient than DL-based approaches since it utilizes a very short duration of the acquired vibration signal (only 0.1 seconds) and produces six features that are sufficient to carry the tasks of fault detection and diagnosis. On the other hand, DL-based approaches usually require extensive training and result in substantial sizes of deep-learned features that necessitate the use of dimensionality reduction feature ranking techniques, which would further increase the computational burden.

\section{Conclusion}
In this paper, a new method is proposed for vibration-based condition monitoring of rolling bearings. The proposed method effectively utilizes instantaneous frequency along with the envelope of generated vibration patterns to jointly facilitate three novel envelope representations: instantaneous amplitude-frequency mapping (IAFM), instantaneous amplitude-frequency correlation  (IAFC), and instantaneous energy-frequency distribution (IEFD). The introduced representations effectively capture unique variations in energy and frequency between healthy and faulty bearings, thereby enabling the extraction of discriminative features with high sensitivity to variations in operational conditions. Accordingly, six new highly discriminative features are extracted from these representations. The experimental results demonstrated excellent performance in detecting and diagnosing various fault types, marking the effectiveness of the proposed method in capturing unique variations in energy and frequency between healthy and faulty bearings. Moreover, the proposed method has comparable performance to DL-based methods but with more moderate computational requirements attributed to the short duration of the utilized vibration segments, efficient feature extraction, and the small set of resulting features.

\bibliographystyle{IEEEtran} 

\bibliography{References}{}

\end{document}